\documentclass[showpacs,preprintnumbers,amsmath,amssymb,nofootinbib]{revtex4-2}

\usepackage{graphicx}
\usepackage{dcolumn}
\usepackage{bm}
\usepackage{color}
\usepackage[normalem]{ulem}
\usepackage{hyperref}

\newcommand{\bq}{\begin{equation}}
	\newcommand{\eq}{\end{equation}}
\newcommand{\bqa}{\begin{eqnarray}}
	\newcommand{\eqa}{\end{eqnarray}}

\newcommand\anote[1]{\textcolor{magenta}{\bf \,#1}}

\def\bnn    {\begin{eqnarray*}}
	\def\enn    {\end{eqnarray*}}

\def\cN {\mathcal{N}}
\begin{document}
	
	\title{Global shift symmetry on an ADM hypersurface: Toward emergent gravity} 
	\author{Ki-Seok Kim$^{a,b}$, Arpita Mitra$^{a}$, Debangshu Mukherjee$^{b}$ and Mitsuhiro Nishida$^{a}$}
	\affiliation{$^{a}$Department of Physics, POSTECH, Pohang, Gyeongbuk 37673, Korea \\ 
		$^{b}$Asia Pacific Center for Theoretical Physics (APCTP), Pohang, Gyeongbuk 37673, Korea}
	
	\email[(Corresponding author) Arpita Mitra: ]{arpitamitra89@gmail.com}
	\email[Debangshu Mukherjee:]{debangshu.mukherjee@apctp.org}
	\email[Ki-Seok Kim: ]{tkfkd@postech.ac.kr}
	\email[Mitsuhiro Nishida: ]{mnishida124@gmail.com}

	\begin{abstract}
		Generalized symmetries and their spontaneous breakdown serve as the fundamental concept to constrain the many-body entanglement structure, which allows us to characterize quantum phases of matter and emergent collective excitations. For example, emergent photons may be understood by spontaneous 1-form symmetry breaking, which results from a long-ranged entanglement structure between UV microscopic degrees of freedom. In this study, we show that emergent ``gravity" may also arise in a similar fashion, where quotes have been used to emphasize that the symmetry-constrained gravitons show unconventional properties compared to usual gravitons. As the electric 1-form symmetry in Maxwell theory is realized as a global shift symmetry of the spatial component of the U(1) gauge field, generated by the electric field, we demonstrate that a constant shift of the Arnowitt-Deser-Misner (ADM) metric on the spatial hypersurface can be viewed as a global symmetry, generated by the ADM canonical momentum. Deriving a vector-type conserved charge from the variation of action, we construct a shift symmetry operator. Considering a Wick rotation, we demonstrate that a gravitational Wilson loop is charged under the action of this shift symmetry operator, which thus confirms the existence of a generalized global symmetry on the ADM hypersurface. Based on the Ward identity, we show that the spontaneous breaking of this global shift symmetry may give rise to a nonpropagating massless symmetric gauge field at the boundary of the hypersurface. 
	\end{abstract}
	
	\maketitle
	\section{Introduction}
	It is a long-standing puzzle to figure out how long-ranged entanglement structure between ultraviolet (UV) microscopic degrees of freedom appears at low energies due to their strong correlations, where artificial photons emerge as collective excitations \cite{Savary:2016ksw,Balents:2010wrb,1905.07040}. In fact, several lattice models allow such deconfined elementary excitations at low energies \cite{Savary:2016ksw,Balents:2010wrb,1905.07040,Nussinov:2009zz,Wen:2018zux,Kobayashi:2018yuk}. Although it is not easy to understand the dynamic process from UV to infrared (IR), one can describe this exotic IR physics from symmetry perspectives \cite{Nussinov:2009zz}, referred to as generalized symmetries and their spontaneous breaking \cite{Gaiotto:2014kfa}. It turns out that pure Maxwell theory without matter has both electric and magnetic 1-form symmetries, and photons arise as Goldstone bosons from the electric 1-form symmetry breaking \cite{Gaiotto:2014kfa,Bhardwaj:2023kri}. Here, Maxwell's equations themselves play the role of current conservation equations, which in turn give rise to conserved charges corresponding to the 1-form symmetries. Moreover, this 1-form symmetry can be visualized with the introduction of an extended object, a string field in $(3+1)$ dimensions (one time and three space dimensions), referred to as vortex string field theory, which occurs from the boson-vortex duality transformation for the Abelian-Higgs model in $(3+1)$ dimensions \cite{Rey:1989ti,Franz:2006gb,Beekman:2010zx,Iqbal:2021rkn}. Indeed, such photons are given by the condensation of vortex string fields, which corresponds to the 1-form symmetry breaking in the conventional sense.
	
	More interestingly, this 1-form symmetry has been used to distinguish two types of superconductors in quantum chromodynamics (QCD) \cite{Hirono:2018fjr,Hirono:2019oup}: (i) nucleon superconductors at low chemical potentials and (ii) quark superconductors at high chemical potentials \cite{Fukushima:2010bq,Schafer:1998ef}. Both superconducting phases exhibit the Higgs phenomenon, where U(1) electromagnetic fields become massive. However, the superconducting state from the condensation of quark Bardeen-Cooper-Schrieffer pairs shows the $\mathbb{Z}_{3}$ center symmetry breaking, which results from the condensation of the Polyakov loop order parameter \cite{Hirono:2018fjr,Hirono:2019oup}. This discrete {1-form} symmetry is preserved in the nucleon superconducting phase due to confinement. As a result, it was claimed that these two superconducting states cannot be connected adiabatically at zero temperature. This discrete {1-form} symmetry breaking perspective has been applied to high T$_c$ cuprates \cite{Kim:2019qtb}. Here, the underdoped superconducting phase is assumed to originate from deconfined fractionalized degrees of freedom, referred to as a spin-liquid-type normal state. On the other hand, the overdoped superconducting state is suggested to result from electron quasiparticle excitations, referred to as {Landau's} Fermi-liquid normal state. Since the former superconducting state shows a discrete {1-form} symmetry breaking in contrast to the other, these two superconducting phases cannot be adiabatically connected to each other at zero temperature inside the superconducting dome of the high T$_c$ cuprates phase diagram.
	
	Now, the question is how we can describe emergent gravity from the perspective of spontaneous breaking of global symmetries, which sometimes occurs in {low-energy} projective Hilbert spaces of certain lattice models, at least in the linearized gravity approximation \cite{Kleinert:2003za,cond-mat/0609595,Xu:2010eg,Pretko:2017fbf}. Although the Weinberg-Witten theorem states that massless particles with spin greater than one cannot carry a Lorentz-covariant stress-energy tensor \cite{Weinberg:1980kq}. This implies that one cannot construct a relativistic QFT with an emergent graviton in the infrared. However, gravity can appear as a gapless mode in effective field theories enjoying biform symmetries \cite{Hinterbichler:2022agn}, which are essentially a subclass of generalized symmetries. \cite{Benedetti:2022zbb}, explores Noether's theorem for generalized symmetries in the context of relativistic QFT--a global continuous generalized symmetry is associated with a local conserved current. The conserved current can be further used to construct the topological charge associated with the corresponding generalized symmetry. Similar to the ordinary symmetries spontaneous breakdown of continuous generalized symmetries can give rise to Goldstone modes \cite{Lake:2018dqm}. 
	
	In this study, we share essentially the same perspective for emergent gravity similar to the electric 1-form symmetry and its spontaneous breaking for emergent photons. We first revisit how electric 1-form symmetry in Maxwell's theory is realized as a global shift symmetry of the spatial component of the $U(1)$ gauge field, generated by the electric field, which is the canonical conjugate of the spatial component of the $U(1)$ gauge field \cite{Gomes:2023ahz,Bhardwaj:2023kri}. We recall that the electric field serves as the conserved charge for the electric 1-form symmetry, where the corresponding conservation law is given by the Maxwell equation itself. In a similar spirit to Maxwell's theory, we make a modest attempt to understand the implications of a global shift in gravity from the canonical quantization perspective. It is well known that perturbative quantization techniques applied to small perturbations around a Minkowski background {fail} after one loop \cite{tHooft:1974toh,Goroff:1985sz}. On the contrary, one can consider a nonperturbative approach via canonical quantization on a constant time slice \footnote{Precisely speaking, it lacks a notion of time evolution in contrast to other canonically quantized theories since the Hamiltonian acts as a constraint of the theory.}.
	
	Working in the ADM formalism \cite{Arnowitt:1959ah} with the specific gauge choice where the lapse and shift functions are constants \footnote{Due to the choice of a hypersurface foliation, general covariance is explicitly broken in the resulting expression and we are left with quantities defined with respect to the induced spatial metric on the hypersurface.}, we determine the conserved generalized Noether current (symmetric rank-2 tensor) associated with the shift of induced metric on the hypersurface and construct the conserved charge. 
	Here, we demonstrate that a constant-in-time shift of the ADM metric can be viewed as a global symmetry on the ADM spatial hypersurface, generated by the ADM canonical momentum. More precisely, we show that this symmetry generates a conserved vector charge on the boundary of the ADM hypersurface. Subsequently, we construct a shift symmetry operator by exponentiation of the vector charge. Compared to the {higher-form} symmetries, here the rank-2 current is symmetric in nature. One can identify the vector charge as conserved linear momentum associated with spatial translation at infinity. Similar vector charges are also observed arising in the rank-2 symmetric $U(1)$ vector charge theory \cite{Pretko:2016kxt, Nandkishore:2018sel}. 
	
	One of the challenges for a quantum gravity theory is to construct a local gauge invariant observable {that} commutes with all the constraints. One can introduce nonlocal observables {starting} from an asymptotic boundary, where all pure diffeomorphisms vanish. The line operator corresponds to a worldline of a probe particle \cite{Donnelly:2015hta,Alfonsi:2020lub}. Since the charge is defined at spatial infinity where no probe particle asymptotes, we consider a Wick rotation. We construct a {line operator} associated with a massless probe and demonstrate that it is charged under the global shift symmetry of the induced metric on the ADM hypersurface. This confirms the existence of a generalized global symmetry on the ADM hypersurface. As a result, we obtain an explicit form of the Ward identity for the ADM shift symmetry. Based on this Ward identity, we figure out {the} spontaneous breaking of this global shift symmetry and the Goldstone theorem by working in momentum space. We demonstrate an explicit pole structure which may imply the presence of nonpropagating massless excitations in the spectrum \cite{QFT_textbook} \footnote{Hence, one can treat the {vacuum} expectation value of this charged object as an order parameter for the braking of generalized shift symmetry.}. This demonstration gives a concrete picture of the generalized global shift symmetry for the gravity theory on a spatial slice in the infrared limit. We point out that electromagnetic formulation of linearized gravity theories and {nonlocal} operators associated with generalized symmetries of gravitons have been also considered recently in Ref. \cite{Benedetti:2021lxj}. Here, in the weak field limit a set of topological charges are constructed out of electric and magnetic components of the linearized Riemann tensor.
	
	\section{Review of shift symmetry in electromagnetism}\label{sec:2}
	First, we briefly review the 1-form symmetry \cite{Gaiotto:2014kfa,Iqbal:2021rkn} of electromagnetism from the perspective of canonical quantization \cite{Gomes:2023ahz}. Pure $U(1)$ gauge theory has two global 1-form symmetries: electric symmetry and magnetic symmetry. Electric 1-form symmetry induces a time-independent shift in the spatial component of the gauge field. To understand this more concretely, we begin with the Lorentz invariant Maxwell Lagrangian density in four spacetime dimensions,
	\begin{equation}
		{\cal L}=-\frac{1}{4}F_{\mu\nu}F^{\mu\nu} .
	\end{equation} 
	
	We consider the field theory to be on a constant time slice with an inertial frame coordinate ($t, x^i$). We also work in the Coulomb gauge given by $A_0=0$ and $\nabla_iA^i=0$ in the absence of any charges. Then, one can construct the phase space in terms of conjugate variables $A^i$ and its canonical momenta, 
	\begin{align}
		\pi^i=\frac{\partial{\cal L}}{\partial\dot{A}_i}=-E^i\label{EB} .
	\end{align}
	The conjugate momentum corresponding to the temporal component, $\pi^0=\frac{\partial{\cal L}}{\partial\dot{A}_0}=0$ is a primary constraint of the theory. In addition, we have a Gauss law constraint that has to be imposed to retain the states invariant under gauge transformation. Imposing this constraint as the selection rule for physical states, we write the commutation relation between the conjugate variables as
	\begin{equation}
		[A^i(\vec{x}), E^j(\vec{y})]=-i\delta^{ij}\delta^{(3)}(\vec{x}-\vec{y})\ .
	\end{equation}
	Under a constant shift of the spatial component of gauge field, $A_i\rightarrow A_i+\lambda_i$, the action transforms as
	
	\begin{equation}\label{maxshift}
		\delta S_{EM}=\int dt\ d^3x \left(\partial_t F^{0i}+\partial_j{F}^{ji}\right)\lambda_i\equiv \int dt\ d^3x \partial_{\mu}J^{\mu i}\lambda_i\ .\qquad (\mu=0,j)
	\end{equation}
	Thus, shifting $A_i$ with a flat connection $\lambda_{i}$ (closed but not exact) is a {global} symmetry with a conserved 2-form antisymmetric current $J^{\mu\nu}=F^{\mu\nu}$. The flatness condition is required since we are interested in global symmetries under which the Wilson loop gains a phase while for gauge symmetries it remains invariant. This kind of transformation is classified by
	the first cohomology group of the spacetime manifold \cite{Gaiotto:2014kfa}.
	%
	%
	Another important point to remember is that the associated current has to be gauge invariant so that one can have a well-defined charge operator in the theory. Interestingly, here the 2-form current $J^{\mu\nu} \equiv F^{\mu \nu}$ is already gauge invariant. 
	
	The corresponding conserved charge is an electric charge computed from the integration over a closed two-dimensional spatial submanifold $\Sigma_2$,
	\begin{align}
		Q_e^{i}=\int_{\Sigma_{2}}d{\Sigma_{2}}^{(i)}~ E^i\ .
	\end{align}
	Therefore, we can define the unitary operator associated with shift symmetry, acting on the Hilbert space in the following way
	\begin{align}
		U_e = e^{i \lambda_{i} Q_{e}^{i}} = e^{i \lambda_{i} \int_{\Sigma_{2}}d{\Sigma_{2}}^{(i)}~ E^i}\ ,\label{EMUO}
	\end{align}
	where $\lambda_{i}$ is the parameter associated with the above transformation. 
	
	The existence of any quantum mechanical symmetry implies that the unitary operator associated with it has to commute with the Hamiltonian of the theory and also has to satisfy the group multiplication rule \cite{Gaiotto:2014kfa,Benedetti:2022zbb}. The operator $U_e$ as defined above in \eqref{EMUO} satisfies all of these properties. As mentioned before, generalized symmetries act on nonlocal objects, in particular a 1-form symmetry will act on line objects.  Under the action of \eqref{EMUO}, the Wilson line associated with a given contour $C$,
	\begin{equation}\label{WLEM}
		W_e[C]=\exp\left[-iq\oint_{C}dy^{i}A_{i}\right]\ , 
	\end{equation}
	transforms as
	\begin{align}
		W'_e=U_eW_eU_e^{\dagger} \ .
	\end{align}
	Using the Baker–Campbell–Hausdorff (BCH) formula, this expression can be simplified to write
	\begin{align}
		W'_e=&\ e^{i \lambda_{i} \int_{\Sigma_{2}}d{\Sigma_{2}}^{(i)}~E^i(\vec{x})}\ e^{-iq\oint_{C}dy^{j}A_{j}(\vec{y})}\ e^{-i \lambda_{i} \int_{\Sigma_{2}}d{\Sigma_{2}}^{(i)} E^i(\vec{x})}\notag\\
		=&\ W_e~e^{i q \int_{\Sigma_{2}}(d{\Sigma_{2}})^{(i)}\oint_{C} d y^{i} \lambda_{i} \delta^{(3)}(\vec{x}-\vec{y})} .
	\end{align}
	Thus, one can see that the Wilson line acquires a phase when we move the 1-form symmetry generator across it, where the infinitesimal change in the Wilson line is,
	\begin{equation}
		\delta W_e[C]=i q \int_{\Sigma_{2}}(d{\Sigma_{2}})^{(i)}\oint_{C} d y^{i} \lambda_{i} \delta^{(3)}(\vec{x}-\vec{y}) W_e[C] = i q \oint_{C} dy^{\nu} \xi_{\nu} W_e[C]\ .
	\end{equation}
	The last equality is expressed in a covariant manner, where $\xi_{\nu} = \delta_{\nu i} \int_{\Sigma_{2}}(d{\Sigma_{2}})^{(i)} \lambda_{i} \delta^{(3)}(\vec{x}-\vec{y})$. In an analogous fashion the magnetic 1-form symmetry can be observed as a shift symmetry of the spatial component of the dual gauge field in the dual formulation of Maxwell's theory \cite{Gaiotto:2014kfa,Iqbal:2021rkn}.
	
	We can understand the conservation law under the symmetry quantum mechanically from the invariance of correlation functions of charged objects, which leads to the so-called Ward identity \cite{Gomes:2023ahz}. {
		If we demand that the expectation value of the charge operator given in \eqref{WLEM} is invariant under the infinitesimal transformation, we require
		\begin{equation}\label{eq:EMward1}
			\langle W_{e}[C]\rangle =\int \mathcal{D}A_{\mu} W_{e}[C] e^{iS_{\text{EM}}[A_{\mu}]}=\int \mathcal{D} A_{\mu}' W'_{e}[C]e^{iS_{\text{EM}}[A_{\mu}']}
		\end{equation}
		where we can define
		\begin{equation}
			\int \mathcal{D} A_{\mu}' W'_{e}[C]e^{iS_{\text{EM}}[A_{\mu}']}=\int \mathcal{D}A_{\mu} \left(W_{e}[C]+\delta W_{e}[C] \right)(1+i \delta S_{\text{EM}})e^{iS_{\text{EM}}[A_{\mu}]}
		\end{equation}
		Eq. \eqref{eq:EMward1} eventually leads us to the following Ward identity} 
	\begin{align}
		i\langle \delta S_{\rm EM} ~ W_e[C]\rangle=-\langle \delta W_e[C]\rangle\ .
	\end{align}
	Plugging \eqref{maxshift} into the above, we can simplify to obtain
	\begin{align}
		\langle \partial_{\mu}J^{\mu\nu}(x) W_e[C]\rangle = - q \int_C dy^{\nu} \delta^{(4)} (x-y)\langle W_e[C]\rangle\label{WT}\ .
	\end{align} 
	Carrying out the Fourier transformation on both sides, we obtain 
	\begin{align}
		i p_{\mu} \langle J^{\mu\nu}(p) W_e[C]\rangle = q \int_C dy^{\nu} e^{ipy}\langle W_e[C]\rangle\label{WT1}\ .
	\end{align}
	
	An important role in the study of low-energy effective field theories is played by the Goldstone theorem, which states that whenever a global symmetry is spontaneously broken, a gapless mode will appear \cite{QFT_textbook}. Similarly, a broken continuous higher-form symmetry in the presence of an extended object leads to a massless Nambu-Goldstone boson \cite{Gaiotto:2014kfa, Lake:2018dqm}. A spontaneous symmetry breaking implies a nonvanishing vacuum expectation value of the charged object, which acts as an order parameter. With $\langle W_e[C]\rangle \neq 0$, one can realize from \eqref{WT1} that the left hand side admits a pole in the vanishing $p$ limit as follows
	\begin{equation}
		\langle J^{\mu\nu}(p) W_e[C]\rangle \propto \frac{p^{\mu}f^{\nu}-p^{\nu}f^{\mu}}{p^2}\langle W_e[C]\rangle ,
	\end{equation}
	where $f^{\nu}=\int_C dy^{\nu} e^{ipy}$ which is nonzero in the vanishing $p$ limit and satisfies $p_{\nu}f^{\nu}=0$ for any closed loop. Here, we get a photon in the Coulomb phase, which indicates the emergence of gauge symmetry from the spontaneous breaking of the global 1-form electric symmetry. In the next section, we will study a similar shift symmetry for gravity theories with ADM foliation.
	\section{Shift symmetry of ADM metric on the hypersurface}
	Since the 1-form symmetry of Maxwell electrodynamics can be interpreted as a shift of the gauge field in the canonical quantization framework, here we will investigate the shift of metric in four spacetime dimensions within the Hamiltonian formulation which inherently necessitates the foliation of spacetime using the so-called ADM formalism \cite{Arnowitt:1959ah}. In this formulation, the spacetime metric is split into spatial and temporal components, and the whole spacetime is now described by hypersurfaces at constant time propagating forward with time. More precisely, ADM formulation is a framework for describing the dynamics of general relativity in terms of induced metrics on a constant time slice and their derivatives. A set of variables is introduced that makes it easier to solve the equations of motion with a given choice of time. These variables include the lapse function $N$, the shift vector $\cN^a$, and the induced metric $h_{ab}$ on the hypersurface. The lapse function and shift vector parametrize the time evolution of the hypersurface.  Canonical quantization of gravity \cite{DeWitt:1967yk} is more natural in such a $3+1$ decomposition of spacetime and has been used extensively in the literature. It is thus convenient to decompose the (3+1) D spacetime metric in terms of ADM variables $N, \cN^a, h_{ab}$ as
	
	\begin{equation}
		ds^2=-N^2 dt^2+ h_{ab}(dx^a +\cN^a dt)(dx^b +\cN^b dt) ,
	\end{equation}
	where the Latin indices $a,b$ run over all spatial indices\footnote{Useful relations are discussed in Appendix \ref{App1}}. 
	
	A commonly adopted gauge choice is $N=1$ and $\mathcal{N}^a=0$, while allowing $h_{ab}$ to vary with time. This choice results in a foliation using Gaussian normal coordinates \cite{Gourgoulhon:2007ue}, frequently employed in the context of cosmological spacetimes, {written explicitly as}
	\begin{equation}\label{metricansatz}
		ds^2= -dt^2+ h_{ab}(\vec{x},t)dx^a dx^b .
	\end{equation}
	We will use this gauge choice to investigate the shift symmetry of the induced metric. 
	Varying the action \eqref{secondform} with respect to the induced metric, and plugging in the expression of the conjugate momentum, we obtain the equation of motion 
	\begin{equation}
		G^{ab}=\dot{\pi}^{ab}+\sqrt{h}[\bar{R}^{ab}-\frac{1}{2}\bar{R} h^{ab}]
		+\frac{1}{\sqrt{h}}\left(2\pi^a_{\,\,c}\pi^{cb}-\pi \pi^{ab}-\frac{1}{2}\left(\pi_{cd}\pi^{cd}-\frac{1}{2}\pi^2\right)h^{ab}\right)=0\ ,
	\end{equation}
	where $G_{ab}$ is the spatial component of the four dimensional Einstein tensor.\footnote{{The other components, namely the $G^{00}$ and $G^{0a}$ components can be thought of as the Hamiltonian and momentum constraints.}}
	Under a variation of {the} induced metric, the ADM action on the hypersurface with the presence of a boundary transforms as \cite{Regge:1974zd},
	\begin{align}\label{Eq:Variation}
		\delta S_{\rm ADM}=\int dt\ d^3x \left(G^{ab} \delta h_{ab}+{d\over dt}(\pi^{ab}\delta h_{ab})\right) .   
	\end{align}
	The second term arises as a consequence of presence of the boundary. Under a time independent constant shift 
	\begin{equation}
		\delta h_{ab}=\Lambda_{ab} ,
	\end{equation}
	Eq. \eqref{Eq:Variation} simplifies to,
	\begin{equation}\label{Eq:Variation2}
		\delta S_{\rm ADM}=\int dt\ d^3x ~\left(G^{ab}\Lambda_{ab}+\dot{\pi}^{ab}\Lambda_{ab}\right) .
	\end{equation}
	
	To interpret shift as a global generalized symmetry in similiar spirit to the Maxwell case, one can try to visualize the equation of motion as a current conservation equation (at least at a linearized level),
	\begin{align}
		G^{ab}=\partial_{\mu}{\cal J}^{\mu a|b}\qquad (\mu=t, c) 
	\end{align}
	The current ${\cal J}^{\mu a|b}$ is a $(2|1)$ tensor associated with $(1|1)$ biform symmetry \cite{Hinterbichler:2022agn}. However, ${\cal J}^{ca|b}$ involves connection terms that are not invariant under diffeomorphism and therefore cannot be treated as {a gauge invariant} operator acting on the Hilbert space of the theory. On the other hand, a gauge invariant current can be constructed from the boundary contribution in \eqref{Eq:Variation2}, which is one main result of this study .
	
	\cite{Hinterbichler:2022agn} demonstrates that the Fierz-Pauli action at linearized level for the covariant gravity theory has global shift symmetry in addition to linearized diffeomorphism. To preserve shift as a global generalized symmetry one needs to ensure that the Riemann tensor of $\Lambda$ has to be vanishing. This is analogous to the Maxwell case where $\lambda$ in \eqref{maxshift} has to be a flat connection. This also ensures that the nonlocal object charged under these symmetries remains invariant up to a phase. Provided the vanishing Riemann tensor condition, one can consider a special choice for $\Lambda_{\alpha\beta}$ for the shift as \cite{Hinterbichler:2022agn},
	
	\begin{equation}
		\Lambda_{\alpha\beta}=\partial^{\rho}\Lambda_{\rho \alpha|\beta}+\partial^{\rho}\Lambda_{\rho\beta|\alpha} .
	\end{equation}
	The corresponding symmetry for the complete covariant theory is well understood in the linearized first order Palatini formulation, where one treats the metric and connection both as independent variables. In this formulation, the linearized gravity theory admits two higher rank currents $J_{\mu\nu|\alpha}$ and $J_{\mu\nu|\alpha\beta}$ associated with $h_{\mu\nu}$ and $\Gamma_{\mu\nu|\rho}$ (symmetric in $\mu\nu$). The $(2|2)$ biform current is gauge invariant, proportional to the Riemann tensor and thus, proportional to second order derivatives of the metric. Imposing the condition of vanishing torsion, the non-gauge-invariant $(2|1)$ current vanishes, and thus, one can identify the $(2|2)$ current associated with the global biform symmetries. This higher-biform symmetry can be gauged by introducing a background gauge field $A_{\mu\nu|\alpha\beta}$. Note that to carry out the analysis described above, \cite{Hinterbichler:2022agn}, did not require any ADM decomposition of spacetime and was able to demonstrate the results respecting covariance.
	
	In contrast to \cite{Hinterbichler:2022agn},	our primary goal is to construct a tensorial current from the boundary contribution of \eqref{Eq:Variation}. Preserving the vanishing Riemann tensor condition, we can as well choose the constant $\Lambda$ as a symmetric combination of the first order derivative of 1-form $\xi$ as
	
	\begin{align}\label{lambdaab}
		\Lambda_{ab}=\partial_a\xi_b+\partial_b\xi_a\ .
	\end{align}
	where the dependence of $\xi$ on spatial coordinates is constrained by the requirement of $\Lambda$ being a constant {or in other words, $\xi_a$ must be linear in coordinates of the hypersurface}. Under this parametrization of shift, the boundary contribution from the variation of action after imposing the equation of motion {boils} down to,
	
	\begin{equation}
		\delta S_{\rm ADM}=-2\int dt\ d^3x \partial_a\left(\partial_t \pi^{ab}\right)\xi_{b}=\int dt\ d^3x~ \partial_{t}J^{0 b}\xi_{b}\label{varac} ,
	\end{equation}
	One can think {of} $J$ as a rank 2 symmetric tensorial current having only the $J^{0b}$ component. Similar tensorial currents also arise as generalized Noether currents in a rank-2 ``{tensor gauge theory}'' \cite{Nandkishore:2018sel}. Note that this expression looks like a variation of an action under a 1-form symmetry. However, compared to a 2-form antisymmetric current for the 1-form symmetry, here the rank 2 tensor is symmetric in nature. 
	
	Now we can construct  a vector-type conserved charge from the integration of $J^{0b}=-2\partial_a{\pi}^{ab}$ on a spatial volume, \footnote{Note that following \cite{Regge:1974zd}, the charge $Q^b$ is essentially a two-dimensional surface integral (performed on the asymptotic region where $r \to \infty$). Assuming that the diffeomorphisms die out ``sufficiently nicely" at this region ensures gauge invariance of the defined charge. {If one computes the charge defined above at the linearized level, following an analysis analogous to \cite{Benedetti:2021lxj}, we presume it will coincide with the Poincar\'{e} charge associated with translations at spatial infinity.}}
	
	\begin{align}
		\label{charge}
		Q^b=-2\int_{\Sigma} d^3x~\partial_a{\pi}^{ab}=-2\int_{\Sigma^{(2)}} d^2x~n_a{\pi}^{ab}=-2\int d\Sigma^{(2)}_a {\pi}^{ab} ,
	\end{align}
	%
	%
	where $\Sigma$ is the ADM hypersurface with a constant time and the boundary $\Sigma^{(2)}$, $n_a$ is unit normal to the boundary. 
	Thus, the vector charge \footnote{One such example includes tensor gauge theories with fractonic symmetries. Similar to (\ref{charge}), an analogous {vector charge} has been considered for theories with fracton quasi-particles, where we have a symmetric rank-2 gauge field with the following gauge symmetry \cite{Benedetti:2021lxj,Nandkishore:2018sel}
		\begin{equation}
			A_{ij}\rightarrow A_{ij}+\partial_i \xi_j+\partial_j \xi_i .
		\end{equation}
		This particular fractonic model is known as a \emph{vector charge tensor model}. One has to impose a generalized Gauss law constraint,
		\begin{equation}
			\partial_i E^{ij}=\rho^j ,
		\end{equation}
		to quantize the theory. Here, charge and dipole moment conservations generate the fractonic gauge symmetries. For a particular choice $\xi_i={1\over 2}\partial_i\xi$, the gauge theory can be shown to be invariant under the transformation $A_{ij}\rightarrow A_{ij}+\partial_i\partial_j \xi$, which is a symmetry for a \emph{scalar charge theory} \cite{Blasi:2022mbl}. } associated with the shift is defined at the boundary of the ADM hypersurface. In addition, it can be identified with the total linear momentum of the system, which is a conserved quantity associated with a global symmetry known as ``spatial translation at infinity'' \cite{Regge:1974zd}. In other words, a particular choice of the shift in the induced metric as \eqref{lambdaab} can be visualized as the generalized global symmetry generated by the spatial translation at infinity. In general, conserved charges associated with higher-form symmetries are constructed by integrating out the Hodge dual of the current, and thus they are defined on subdimensional spaces. These charges are thus, by construction topological in nature. The shift charge in \eqref{charge} can be used to construct a topological operator, which we now discuss below. 
	
	Next, we consider how this charge acts on a nonlocal object. We construct the shift symmetry operator, exponentiating the conserved and gauge-invariant charge defined in \eqref{charge} as follows

	\begin{equation}
		U_h=\exp\left(2i\xi_b\int_{\Sigma^{(2)}}  d^2 x~ n_a\pi^{ab}\right)\label{sym} .
	\end{equation}
	where $\xi_b$ is the parameter of the shift symmetry on the boundary. $U_h$ is an extended operator with support on a codimension 2 surface ${\Sigma^{(2)}} $ at the spatial boundary of the spacetime. Any global symmetry is known to be associated with a topological operator since the charge is conserved under any deformation of {the} smooth manifold on which it is defined \cite{Gaiotto:2014kfa}. For gravity theories with  ADM foliation all spacelike Cauchy slices  (globally defined) asymptote to the spatial infinity where $U_h$ is defined. Thus $U_h$ remains unchanged under any smooth deformations of the slice and therefore can be thought of as a topological operator.
	
	In canonical quantization framework of gravity on a compact manifold, there exist no local observables which commute with all constraints of the theory. Therefore for gravity theories it is more natural to construct a gauge invariant nonlocal observable. Since generalized symmetries also act only on nonlocal operators such as the Wilson line, we act this shift symmetry operator \eqref{sym} on a gravitational Wilson line. However the charge has support only on the spatial infinity and no massive or massless particles asymptote to the spatial infinity of asymptotically flat spacetimes. Nevertheless spacelike Wilson loops do exist in Euclidean space, and therefore we will incorporate a Wick rotation. 
	
	A gravitational Wilson line can be defined in terms of the phase experienced by a scalar test particle in a gravitational field \cite{Alfonsi:2020lub, Ambjorn:2015rva}. This must be proportional to the (Euclidean) action of a point particle of mass $m$ with spacetime trajectory $C$ in a Riemannian geometry with metric $g_{\mu\nu}^E$, {motivating us to} introduce the phase as
	
	\begin{align}
		\Phi=\left[m\int_{C} ds \left(g_{\mu\nu}^E(X(s))\frac{dX^{\mu}}{ds}\frac{dX^{\nu}}{ds}\right)^{1/2}\right]\label{WL}\ ,
	\end{align}
	where $C$ represents an arbitrary curve parametrized by $X(s)$.\footnote{In the weak field limit, one can consider small perturbations around the Minkowski spacetime,
		\begin{align}
			g_{\mu\nu}=\eta_{\mu\nu}+\kappa~ \gamma_{\mu\nu} .
		\end{align}
		Carrying out an expansion for the gravitational Wilson line up to the linear order of $\kappa$, $\Phi$ can be written as 
		
		\begin{align}
			\Phi=\exp\left[\frac{i\kappa}{2}	\int_C ds \frac{dX^{\mu}}{ds}\frac{dX^{\nu}}{ds}\gamma_{\mu\nu}(x(s))\right] \label{WL_Perturbation} .
		\end{align}
		Here, the mass $m$ was absorbed into the length parameter $s$.}	
	The issue of square root ``singularity" can be avoided with the interpretation of $\Phi$ as the Euclidean action of a relativistic point particle. We can rewrite the point-particle action in the following form \cite{Polchinski},
	
	\begin{align}
		S_{\text{PP}}={1\over 2}\int_C e(s)~ds\left({1\over e^2(s)}g_{\mu\nu}^E\frac{dX^{\mu}}{ds}\frac{dX^{\nu}}{ds}+m^2\right)\ ,
	\end{align}
	where $e(s)$ is an auxiliary field acting as a vielbein on the worldline.   Substituting $e$ with the solution of its equation of motion, one can get back \eqref{WL}. $e(s)$ essentially plays the role of Lagrange multiplier and the action enjoys reparametrization invariance. One can use this symmetry to fix $e(s)=1$ which leads to the following Wilson line operator,
	
	\begin{equation}\label{WLnew}
		W_g=  \exp\left[{i\over 2}\int_{C} ds~\left( g_{\mu\nu} \frac{dX^{\mu}}{ds}\frac{dX^{\nu}}{ds}+m^2\right)\right]\ .  
	\end{equation}
	Considering the ADM foliation and taking the vanishing mass limit, \eqref{WLnew} gives,
	
	\begin{equation}
		W_h=\exp\left[{i\over 2}\int_{C} ds\left(\frac{dy^{0}}{ds}\frac{dy^{0}}{ds}+\frac{dy^{a}}{ds}\frac{dy^{b}}{ds}h_{ab}(\vec{y}(s))\right)\right] \label{line} .
	\end{equation}
	Under the symmetry operation \eqref{sym}, the line operator in \eqref{line} transforms in the following manner, 	
	
	\begin{align}\label{W}
		W_h'=U_h~W_h~ U^{\dagger}_h\ .
	\end{align}
	To compute the above after a Wick rotation $\pi_{ab}\rightarrow i\pi_E^{ab}$, we need to use the Baker-Campbell-Hausdorff formula, which further requires the following commutator	
	
	\begin{align}\label{Eq:intcom}
		\left[-2\xi_{d}\int d^3 x ~\partial_c\pi^{cd}_E(\vec{x}), {i\over 2}\int_C ds \frac{dy^{a}}{ds}\frac{dy^{b}}{ds}h_{ab}^E(\vec{y}(s))\right]&=i\xi_{b}\int d^3 x \int_C
		ds \frac{dy^{a}}{ds}\frac{dy^{b}}{ds}~\partial_{a}\delta^{(3)}(\vec{x}-\vec{y}(s)) .
	\end{align}
	In deriving the above, we have used the commutation relation resulting from the uplifting of Poisson bracket between the conjugate variables \eqref{PB}. Generally speaking, within the canonical quantization framework, after solving the constraint equations one can construct the reduced phase space and then upgrade the Poisson bracket \eqref{PB} to a commutator. After a Wick rotation the commutation relation becomes,
	\begin{equation}\label{ECB}
		[h_{ab}^E(\vec{x},\tau),\pi^{cd}_E(\vec{y},\tau)]=\frac{1}{2}(\delta^c_a\delta^d_b+\delta^d_a\delta^c_b){\delta^{(3)}(\vec{x}-\vec{y})} .
	\end{equation}
	
	Under the action of the shift symmetry operator, we observe that the Wilson line acquires a phase under the shift (similar to the case of free Maxwell theory) as
	
	\begin{equation}
		W_h'=\exp\left(i~ \xi_{b}\int_C ds \frac{dy^{a}}{ds}\frac{dy^{b}}{ds}\chi_a(\vec{y})\right)W_h .
	\end{equation}
	where we have introduced $\chi_a$, following \cite{Gomes:2023ahz} as
	
	\begin{equation}
		\chi_a(\vec{y}) =\int d^3 x~ \partial_{a}\delta^{(3)}(\vec{x}-\vec{y}) = \int d^2 x~ n_{a}\delta^{(3)}(\vec{x}-\vec{y}).
	\end{equation}
	The above confirms that the shift symmetry is indeed a ``global" symmetry for the Wilson line on the ADM hypersurface. 
	
	This generalized global symmetry would give rise to a Ward identity, regarded as a consequence of Noether's theorem at the quantum level. The Ward identity can be computed from the invariance of correlators under this symmetry.  We consider the expectation value of the Wilson line in the path integral framework as\footnote{A careful consideration of the measure requires a gauge fixing by introducing a Fadeev-Poppov determinant.}
	
	\begin{align}
		\langle W_h[C]\rangle \sim \int {\cal D}h_E~ W_h[C] \exp(- S_E[h]) .
	\end{align}
	Ensuring the invariance of the expectation value under the shift of the induced metric (in the absence of anomalies) \cite{Gomes:2023ahz}, we obtain,
	
	\begin{align}
		\langle W_h[C]\rangle=\int {\cal D}h'_E W_h'[C] \exp(-S_E[h'])=\int {\cal D}h_E W_h'[C] (1-\delta S_E)\exp(-S_E[h]) .
	\end{align}
	Here, we expand the transformed Wilson line as follows	
	\begin{align}
		W_h' = \left(1+i~ \xi_{b}\int_C ds \frac{dy^{a}}{ds}\frac{dy^{b}}{ds}\chi_a(\vec{y})\right)W_h .
	\end{align}
	Recalling the variation of action under the shift after a Wick rotation,
	
	\begin{equation}
		\delta S_E=\int d\tau\ d^3x \partial_{\tau}J^{0 b}\xi_{b} , \nonumber
	\end{equation}
	and introducing this expression into $\langle \delta S_E ~ W_h[C]\rangle=\langle \delta W_h[C]\rangle$, we find the Ward identity as follows
	
	\begin{align}\label{WI}
		\left\langle ~\partial_{\tau}J^{0b}(\tau,\vec{x}) ~ W_{h}[C] \right\rangle=\left\langle i\int_C ds~\partial_{a}\delta^{(3)} (\vec{x}-\vec{y})~\frac{dy^{a}}{ds}\frac{dy^{b}}{ds}  ~  W_h[C] \right\rangle\Big|_{\tau=y_0}\ ,
	\end{align}
	
	{The spontaneous} breaking of this generalized global symmetry is described by the condensation of the Wilson line implying a non vanishing vacuum expectation value. The expectation value of {the} Wilson line may be regarded {as} an order parameter for the phase transition induced by the symmetry breaking. 
		The Ward identity when written in momentum space will explicitly demonstrate the appearance of gapless Goldstone modes - a ``Goldstone theorem" in the present context.

		We first perform the Fourier transformation of the rank 2 tensorial current, 
		
		\begin{equation}\label{Fourier}
			\tilde{J}^{0b}(p^0,\vec{p})=\int d\tau~d^3x~ e^{ip^0 \tau+i\vec{p}.\vec{x}}J^{0b}(\tau,\vec{x})\ , 
		\end{equation}
		where for massless particles, energy and momentum are related according to,
		
		\begin{equation}
			p_0^2+h_{ab}p^ap^b=0 
		\end{equation}
		which comes from \eqref{metricansatz} following a Wick rotation. Condensation of the line operator indicates that $\big\langle W_h[C] \big\rangle \not= 0$. Thereafter plugging back the expressions \eqref{Fourier} in \eqref{WI}, one can obtain 
		
		\begin{align}\label{WI2}
			-2 p_{0}\langle \tilde{\pi}^{ab}_E(p) ~ W_{h}[C] \rangle = {\bf{f}}^{ab}(p^0,\vec{p},C)~ \big\langle W_h[C] \big\rangle .
		\end{align}
		where we denote
		
		\begin{equation}
			{\bf{f}}^{ab}=\int_C ds \frac{dy^{a}}{ds}\frac{dy^{b}}{ds}e^{ip^0 \tau+i\vec{p}.\vec{y}} 
		\end{equation}
		Therefore, we can observe the existence of a pole at $p_0\rightarrow 0$ limit according to,
		\begin{equation}
			\langle \tilde{\pi}^{ab}_E(p) W_h[C]\rangle \propto \frac{{\bf{f}}^{ab}}{p_0}\langle W_h[C]\rangle\ ,
		\end{equation}
		This may be regarded as a Goldstone theorem, which implies the existence of gapless immobile particles on the ADM hypersurface. It would be interesting to consider how this Goldstone boson may be related to fractons to represent immobile excitations \cite{Pretko:2017fbf, Nandkishore:2018sel,Pretko:2020cko, Gromov:2022cxa, Banerjee:2022nin}. In general, gapless fracton phases arise in symmetric tensor gauge theories which exhibit duality with quantum theory of elasticity and also have some resemblance to linearized gravity theories. Here, one may think that the Weinberg-Witten theorem disallows the emergence of gravity from a tensor gauge theory. Nevertheless, the theorem relies on two key assumptions: Lorentz invariance and the presence of asymptotic momentum eigenstates for particles. For fractonic theories, Lorentz invariance is broken and due to immobility isolated fractons do not possess a description in terms of momentum eigenstates, making the theorem not readily applicable in this scenario. Interestingly, these two conditions also hold for the ADM case. In a related direction, \cite{Argurio:2021opr} investigates construction of low-energy effective theories characterized by the emergence of symmetries that give rise to fractonic modes from {the} spontaneous breaking of spatial translations, particularly those that exhibit gaplessness. 
		

		%
		%
		
		\section{Conclusion}
		

		{This study originated from the observation that {0 and 1-form} symmetries {can also be studied from} the commutation relation between the canonical conjugate pairs, where the canonical momentum plays the role of a generator of a shift transformation of the conjugate field. Our aim was to investigate if an analogous description exists for gravity within the Hamiltonian formulation. Interestingly, we find a tensorial rank 2 current which is conserved on-shell provided one has ``nice" falloffs for arbitrary diffeomorphism of the boundary i.e. diffeomorphism that does not disturb the asymptotic structure of the spacetime at spatial infinity on any arbitrary Cauchy slice. The associated charge is subsequently used to define an extended topological operator on a codimension-2 surface at the boundary of spatial hypersurface. We subsequently see the effect of this symmetry operator on a nonlocal gauge invariant observable, namely the Wilson line in a theory of Euclidean gravity. {{We have not considered any possible large gauge transformations at the spatial boundary, which {do not} die off in asymptotic regions. Typically, these kinds of diffemorphisms are parametrized by an infinite family of functions that depend on angular coordinates at the boundary. The charges associated with global shift symmetry of the induced metric on the ADM hypersurface are not invariant under such large diffeomorphisms.}}
			
			We emphasize the significant departure of our analysis from the usual 0 and 1-form symmetries that one sees in $U(1)$ gauge theory.
			\begin{itemize}
				\item Recall that, in free $U(1)$ gauge theory, the equation of motion itself is identified as the current conservation equation associated with 1-form symmetry. In contrast to Maxwell electrodynamics, in the context of Einstein gravity in ADM foliation, the extended topological symmetry generator {{is defined at the boundary of the ADM hypersurface}}. 
				\item In Lorentzian spacetimes, we cannot have any physical particles (massless or massive) reaching spatial infinity--which is precisely where we eventually defined the topological symmetry generators. However, they will have nontrivial effect on Wilson lines in Euclidean spacetime where all asymptotic points are essentially spacelike.
			\end{itemize}
			Furthermore, we derived the corresponding Ward identity and showed that there appears a massless localized excitation on the ADM hypersurface when the gravitational Wilson line is condensed. One might be curious as to the applicability and the physical scenario where such operators and their effect on Euclidean Wilson lines might become relevant. The kind of conserved charges and extended operators discussed in this work will presumably play an important role in the physics of thermal systems, which typically requires one to Wick rotate the temporal coordinate and further compactify it on a circle. However, more sharp and concrete statements require more analysis, which we relegate to future work. 
			
			%
			%

			\begin{acknowledgments}
				This work was supported by the Ministry of Education, Science, and Technology (RS-2024-00337134) of the National Research Foundation of Korea (NRF) and by TJ Park Science Fellowship of the POSCO TJ Park Foundation. The work of A. M. is supported by POSTECH BK21 postdoctoral fellowship. The work of D. M. was supported by the \emph{Young Scientist Training (YST) Fellowship} from Asia Pacific Center for Theoretical Physics. A.M. and D.M acknowledge support by the National Research Foundation of Korea (NRF) grant funded by the Korean government (MSIT) (No. 2022R1A2C1003182). M. N. was supported by the Ministry of Education, Science, and Technology (RS-2023-00245035).
			\end{acknowledgments}
			\appendix
			\section{ADM formalism in (3+1) dimensions}\label{App1}
			In ADM formalism, the (3+1)-dimensional metric is foliated according to,
			\begin{equation}
				ds^2=-N^2 dt^2+ h_{ab}(dx^a +\cN^a dt)(dx^b +\cN^b dt) ,
			\end{equation}
			where the Latin indices $a,b$ run over all spatial indices.The inverse metric components are given by
			
			\begin{equation}\label{staticgague}
				g^{tt}=-\frac{1}{N^2}\ ;\quad g^{at}=\frac{\cN^a}{N^2}\ ;\quad g^{ab}=h^{ab}-\frac{1}{N^2}\cN^a \cN^b\ .
			\end{equation}
			
			With this decomposition, the three dimensional hypersurfaces at constant time are associated with an intrinsic curvature given by 3D Riemann tensor and an extrinsic curvature $K_{ab}$ defined as
			\begin{equation}
				K_{ab}=\frac{1}{2 N}(D_a N_b+D_b N_a-\dot{h}_{ab}), \quad K=h^{ab}K_{ab} ,
			\end{equation}
			where $D_a$ is the covariant derivative with respect to the induced spatial metric $h_{ab}$. The extrinsic curvature arises from the embedding of hypersurfaces in 4D spacetimes. As a result, we can rewrite the Einstein-Hilbert action in terms of the ADM variables as follows
			
			\begin{equation}\label{secondform}
				S_{ADM}^{\text{bulk}}= \frac{1}{\kappa}\int dt\ d^3x \sqrt{h}N(\bar{R}+K^{ab}K_{ab}-K^2) ,
			\end{equation}
			where $\bar{R}$ is the Ricci scalar of the hypersurface {and} $\kappa=16 \pi G_N$ and will set to be unity in the future onward. 
			This ADM action can be also expressed in terms of the DeWitt metric,
			\begin{equation}
				G_{abcd}=\frac{1}{2\sqrt{h}}(h_{ac}h_{bd}+h_{ad}h_{bc}-h_{ab}h_{cd})\ ,
			\end{equation}
			as {follows}
			\begin{equation}
				\begin{aligned}
					S_{ADM}^{\text{bulk}}=\frac{1}{\kappa}\int dt\ d^3x\ \left[N\sqrt{h}(G^{abcd}K_{ab}K_{cd}+\bar{R})\right] .
				\end{aligned}
			\end{equation}
			
			The Hamiltonian {can be} derived by performing the Legendre transformation of the ADM action. This transformation requires a set of conjugate momenta {$\pi^{ab}$} that describes how the spatial metric changes over time and is given by
			\begin{equation}
				\pi^{ab}=\frac{\delta L}{\delta \dot{h}_{ab}}=-\sqrt{h}(K^{ab}-h^{ab}K) .
			\end{equation}
			The phase space will be described in terms of conjugate variables ($h_{ab}, \pi^{ab}$). Consequently, we can express the Hamiltonian arising from the action in Eq. \eqref{secondform} as
			\begin{equation}
				H_{ADM}=\int_{\Sigma} d^3 x~ (N {\cal H}+{\cal N}^a {\cal H}_a) ,
			\end{equation}
			where 
			\begin{equation}
				{\cal H}=\sqrt{h}(K^{ab}K_{ab}-K^2-\bar{R})=\frac{1}{\sqrt{h}} G_{abcd}\pi^{ab}\pi^{cd}-\sqrt{h}~\bar{R}\qquad \text{and}\qquad{\cal H}^b=-2 D_a\pi^{ab} .
			\end{equation}
			
			Taking into consideration the definition of the Hamiltonian, we can also {write} the ADM action \eqref{secondform} in first order form as
			\begin{equation}
				S_{ADM}^{\text{bulk}}=\int dt\ d^3x\ \left[\pi^{ab}\dot{h}_{ab}-N\mathcal{H}-\cN^a \mathcal{H}_a\right] ,
			\end{equation}
			which, upon variation with respect to the lapse and shift, yields the well-known Hamiltonian and momentum constraints,
			\begin{equation}
				{\cal H}=0,\qquad {\cal H}_a=0 .
			\end{equation}
			Variation with respect to $\pi^{ab}$ and $h_{ab}$, results in Hamilton's equations of motion,
			\begin{align}
				&\dot{h}_{ab}=\frac{\delta H}{\delta \pi^{ab}}=\{ h_{ab}, H_{\text{ADM}}\},\notag\\& \dot{\pi}^{ab}=-\frac{\delta H}{\delta h_{ab}}=\{ \pi^{ab}, H_{\text{ADM}}\} ,
			\end{align}
			where the equal time Poisson bracket between the canonical variables is given by
			\begin{equation}\label{PB}
				\{h_{ab}(\vec{x}),\pi^{cd}(\vec{y})\}=\frac{1}{2}(\delta^c_a\delta^d_b+\delta^d_a\delta^c_b){\delta^{(3)}(\vec{x}-\vec{y})} .
			\end{equation}
			
			In the presence of a boundary, one has to add a boundary term to the bulk Hamiltonian, 
			\begin{align}
				H_{ADM}^{\rm Bdy}=-\frac{2}{\kappa}\oint d^2x (N\sqrt{s} K_S-{\mathcal{N}}_a \pi^{ab} r_b)\label{HB} ,
			\end{align}
			where $s_{ab}$ is the metric on the sphere at the boundary of the spacelike hypersurface, $K_S$ is the extrinsic curvature of this boundary, and $r^{a}$ is the unit normal to the boundary. One can think of this piece as the on-shell \emph{boundary Hamiltonian} after imposing the Hamiltonian and momentum constraints. The boundary piece in \eqref{HB} is associated with asymptotic time translation and is related to the ADM energy
			\begin{align}
				E_{\text{ADM}}=-\frac{2}{\kappa}\lim_{R\rightarrow \infty}\oint d^2x \sqrt{s} (K_S- K_S^0) 
			\end{align}
			for static spacetimes. Note that, in the above, one has to regularize the ADM energy by subtracting out the ``infinity" coming from the vacuum, denoted by $K_S^0$. 
			\section{Useful quantities in Gaussian normal coordinates}
			In Gaussian normal coordinates, metric, affine connection and Riemann tensor components are given by
			\begin{align}
				& g_{ab}=h_{ab},\quad |g|=|h| , \notag\\
				& \Gamma^{a}_{bc}= \bar{\Gamma}^{a}_{bc},\quad  \Gamma^{0}_{ab}= K_{ab}=-{1\over 2}\dot{h}_{ab},\quad  \Gamma^{a}_{0b}= K^a_{b}, \quad \Gamma^{a}_{00}=\Gamma^{0}_{a0}=\Gamma^{0}_{00}=0,\notag\\
				& R_{abcd}=\bar{R}_{abcd}+(K_{ac}K_{bd}-K_{bc}K_{ad}) , \notag\\&  R_{abc0}=D_aK_{bc}-D_bK_{ac} , \notag\\&  R_{a0b0}=E_{ab}=-\dot{K}_{ac}+{K}_{a}{}^c K_{cb} ,
			\end{align}
			where $E_{ab}$ is the electric component of the Riemann tensor in ADM foliation and $\bar{\{\}}$ denotes three dimensional quantities. 

		\end{document}